%% file: mainEL.tex
\documentclass[11pt]{article}
\usepackage{authblk}
\usepackage{amssymb}
\usepackage{graphicx,color}
\usepackage{amscd}
\usepackage{amsmath}
\usepackage{amsfonts}
\usepackage{mathtools}
\usepackage{fullpage}
\usepackage{tikz}
\usepackage{graphicx}
\usepackage{subcaption}

\newcommand{\ft}{\text}

\newcommand{\ed}{{\text{d}}}

\title{A geometric approach to Hu-Washizu variational principle in nonlinear elasticity}
\author[1]{Bensingh Dhas \thanks{bensingh@iisc.ac.in}}
\author[1,2]{Debasish Roy \thanks{royd@iisc.ac.in}}
\affil[1]{Center of Excellence in Advanced Mechanics of Materials, Indian
Institute of Science, Bangalore 560012, India}
\affil[2]{Computational Mechanics Lab, Department of Civil Engineering, Indian
Institute of Science, Bangalore 560012, India}
\date{}

\begin{document}
\maketitle
\begin{abstract}
    We discuss the Hu-Washizu (HW) variational principle from a geometric
    standpoint.  The mainstay of the present approach is to treat quantities
    defined on the co-tangent bundles of reference and deformed configurations
    as primal. Such a treatment invites compatibility equations so that the base
    space (configurations of the solid body) could be realised as a subset of an
    Euclidean space.  Cartan's method of moving frames and the associated
    structure equations establish this compatibility. Moreover, they permit us
    to write the metric and connection using 1-forms.  With the mathematical
    machinery provided by differentiable manifolds, we rewrite the deformation
    gradient and Cauchy-Green deformation tensor in terms of frame and co-frame
    fields.  The geometric understanding of stress as a co-vector valued 2-form
    fits squarely within our overall program. We also show that for a
    hyperelastic solid, an equation similar to the Doyle-Erciksen formula may be
    written for the co-vector part of the stress 2-form. Using this kinetic and
    kinematic understanding, we rewrite the HW functional in terms of frames and
    differential forms. Finally, we show that the compatibility of deformation,
    constitutive rules and equations of equilibrium are obtainable as
    Euler-Lagrange equations of the HW functional when varied with respect to
    traction 1-forms, deformation 1-forms and the deformation. This new
    perspective that involves the notion of kinematic 	closure precisely
    explicates the necessary geometrical restrictions on the variational
    principle, without which the deformed body may not be realized as a subset
    of the Euclidean space. 	It also provides a pointer to how these
    restrictions could be adjusted within a non-Euclidean setting.  
\end{abstract}
\section{Introduction}
Differential forms provide for a natural descriptor of many phenomena of
interest in science and engineering \cite{flanders1963}. The rules for combining
and manipulating differential forms were developed by Hermann Grassmann; however
it was only in the work of \'Elie Cartan that differential forms found a place
as a tool to study the geometry of differentiable manifolds.  Later, within the
broad program of geometrisation of physics, differential forms became an
indispensable part in the mathematical description of phenomena. One of the
classical examples of a physical theory that was reformulated using differential
forms is that of electro-magnetism, where electric and magnetic fields were
understood as differential forms of degrees $1$ and $2$ respectively
\cite{hehl2012}. This reformulation laid bare the key algebraic and geometric
features of the theory. 

The mechanics of non-linear elastic solids, or the mechanics of continua in
general, has a distinctive difference vis-\'a-vis the theory of
electro-magnetism. To be sure, both the theories have sections from the
co-tangent bundle or the tensor bundle of the co-tangent space as fields of
interest. While electric and magnetic fields are the quantities of interest in
electromagnetism, the interest in elasticity is on stresses and strains. The
distinction arises from the nature of the base space. In the case of
electro-magnetism, the base space is fixed and only sections from the tensor
bundle have an evolution rule. In contrast, for an elastic solid the base
manifold evolves during the deformation process; such evolutions place
restrictions on the evolving sections from the tensor bundle.  This additional
constraint on sections of the tensor bundle can be overlooked if one chooses to
work with deformation as the only variable, when other quantities of interest
are derived from it. Such an approach has to an extent been successful for
elastic solids. But it suffers from serious drawbacks when one tries to extend
the mechanics to inelastic deformation like plasticity, where geometric
assumptions on the configuration of a body are brought in to bear on the
mechanics \cite{sadik2017}.  Cavitation in nonlinear elastic solids
\cite{ball1987} is another classical example where the deformation alone as the
primal variable fails.

It is well understood that the geometry of non-linear elasticity is Eucidean
\cite{mfe}. If one tries to independently evolve stresses and strains along with
deformation, one must as well impose the constraint that the configurations of
the body are subsets of an Euclidean space. We adopt the method of moving frames
proposed by Cartan to achieve this. In this technique, we assign to each point
of the configuration, vectors called frame fields. The rate at which these
frames vary across the configuration defines the connection 1-forms on the
manifold. These connection 1-forms have to satisfy the structure equations so
that the parallel transport they encode conforms to the underlying Euclidean
structure. Attaching a set of vectors to a material point is not new to the
continuum mechanics community. Many such models have been put forth, starting
from Cosserat to micromorphic theories; they are often referred to as
micro-continuum \cite{eringen2012} theories. An important distinction between
the present approach and the so called micro-continuum theories is that the
latter do not use the directors to encode the connection information. For these
models, directors are just degrees of freedom to hold energy. A major trouble
with this point of view is that it does not clarify the geometry within which
the model is working. An immediate consequence is that the it is impossible to
give a co-ordinate independent meaning to the derivatives appearing in the
equations of motion. A similar scenario is also encountered in shell theories.
Often computational schemes for nonlinear shells \cite{simo1989} are built on
models which include directors as degrees of freedom in addition to the
mid-surface deformation. We believe that these director degrees of freedom are
proxies for the connection induced on the surface. Since these directors do not
satisfy any structure equations, it may not be possible to realise these shells
as subsets of an Euclidean space.   

The goal of this article is to reformulate the kinematics and kinetics of
deformation using differential forms. Cartan's method of moving frames is
exploited to construct the kinematics of deformation in terms of frames and
co-frames. The kinematic reformulation we propose is a first of its kind and
hence not available in the literature. In reformulating the kinetics of
deformation, we use the work of Kanso \textit{et al.} \cite{kanso2007}, where
stress was interpreted as a co-vector valued 2-form.  Our kinematic
reformulation is also dual to the understanding of stress as a 2-form proposed
by Kanso \textit{et al.}. Using this kinematic-cum-kinetic formalism, we rewrite
the HW energy functional and show that it recovers the equilibrium equations as
well as compatibility and constitutive rules.

The goals, just stated, have consequences for computation as well. Mixed finite
element methods already discussed in the literature may now be realised as tools
that place geometry and deformation on an equally important footing. The
algebraic and geometric structures brought about by differential forms are
instrumental in this.  Numerical techniques developed in the form of vector
finite elements like the Raviart-Thomas, N\'ed\'elec
\cite{raviart1977,nedelec1980,nedelec1986} and other carefully handcrafted
finite element techniques are now being unified under the common umbrella of
finite element exterior calculus, where differential forms play a significant
role \cite{bossavit1988,arnold2006,arnold2010,hiptmair1999}. Within nonlinear
elasticity, techniques based on mixed methods are already the preferred choice
\cite{shojaei2018,shojaei2019,angoshtari2017} for large deformation problems.
Motivated by the algebra of differential forms, techniques to approximate
differential forms outside the conventional framework of finite elements
\cite{hiraniThesis,yavari2008} are also being explored. These techniques can be
better developed and interpreted using the geometric approach we propose for the
HW functional, when applied to nonlinear elasticity problems.

The rest of the article is organized as follows. A brief introduction to
differential forms is given in Section \ref{sec:diffForms}. Kinematics of an
elastic body is presented in Section \ref{sec:kinematics}, where important
kinematic quantities like deformation gradient and right Cauchy-Green
deformation tensor are reformulated  in terms of frame and co-frame fields. This
reformulation is facilitated by Cartan's method of moving frames. This section
also contains a discussion on affine connections using connection 1-forms. In
Section \ref{sec:stress}, we introduces stress as a co-vector valued
differential 2-form; this interpretation is originally due to Kanso \textit{ et
al.}. Then a result like the Doyle-Ericksen formula is presented which yields a
constitutive description for traction 1-forms in the presence of a stored energy
function.  In Section \ref{sec:variational}, we rewrite the HW variational
principle in terms of differential forms using the kinematics and kinetics of
deformation developed in Sections \ref{sec:kinematics} and \ref{sec:stress}.
Then we show that variations of the HW functional with respect to different
input arguments lead to the compatibility of deformation, constitutive rule and
equations of equilibrium. We also remark on the interpretation of stress as a
Lagrange multiplier enforcing compatibility of deformation. Finally, in Section
\ref{sec:conclusion}, we remark on the usefulness of geometrically reformulated
HW variational principle in constructing efficient numerical schemes for
non-linear elasticity and its extension to other theories in nonlinear solid
mechanics.

\input{dfm}
\input{kinematics}
\input{stress}

\section{Conclusion}\label{sec:conclusion}
We have formulated the HW variational principle using differential forms. The
main tenets of this reformulation are Cartan's method of moving frames and the
interpretation of stress as a co-vector valued 2-form. The Euler-Lagrange
equations clearly explicate how additional stresses could be generated when the
kinematic closure or compatibility conditions are not explicitly imposed. These
stresses are the result of incompatibilities that develop and co-evolve with the
deformation and may render the deformed solid body non-Euclidean. In this sense,
the present variational approach may  be used not only with models that aim at
restricting the deformed body as a subset of the Euclidean space, but also with
those where the evolution of incompatibilities, e.g. defects, is of importance. 

Our novel approach to the  HW variational principle also has consequences in the
numerical solution of the equations of nonlinear elasticity. The discretization
schemes based on finite element exterior calculus may now be seamlessly used to
approximate the differential forms appearing in the HW functional. Such an
approximation has the advantage of respecting the algebraic and geometric
structures defined by these differential forms even after discretization. Work
is currently under way to study such approximations.

The kinematic framework developed to describe deformation has a specific
advantage in modelling the motion of shells. In the case of shells, the
differential of the position vector is non-trivial since the tangent spaces at
different material points are not the same.  When tied with Cartan's moving
frames, the kinematics of the shell surface can be completely reformulated in
terms of differential forms. The stress resultants may then be understood as
bundle valued differential forms.  These kinetic and kinematic ideas may then be
stitched together using the new HW principle, to arrive at a shell theory that
is transparent in its geometric assumptions. 
\section*{Acknowledgements}
BD was supported by ISRO through the Centre of Excellence in Advanced
Mechanics of Materials; grant No. ISRO/DR/0133. 
\newpage
\bibliographystyle{abbrv}
\bibliography{defects}
\end{document}

%% file: dfm
\section{Differential forms in mechanics}\label{sec:diffForms} 
We now provide a brief introduction to differential forms; the material is
standard and can be found in many introductory texts to differential geometry
and manifold theory, e.g.  \cite{tu2010introduction,guggenheimer1963}. Let
$\mathcal{M}$ be a smooth manifold; at any point of $\mathcal{M}$ one may define
the tangent space, which is a vector space. We denote it by $T_x\mathcal{M}$.
The dual space to this vector space is denote by $T^*_x\mathcal{M}$. If $\alpha
\in T^*_x\mathcal{M}$ and $v \in T_x\mathcal{M}$, then the action $\alpha(v)$ is
a real number.  The collection of $T_x\mathcal{M}$ for each $x\in \mathcal M$ is
called the tangent bundle $T\mathcal M :=\bigcup_{x\in \mathcal M}T_x \mathcal
M$; similarly we may define the cotangent bundle as $T^*\mathcal M=
\bigcup_{x\in \mathcal{M}}T^*_x\mathcal M$.  If one picks at each point in
$\mathcal{M}$ an element of $T^*_x\mathcal{M}$, we say that a 1-form is defined
on $\mathcal{M}$. In other words 1-forms are sections from $T^*\mathcal M$, just
as a vector field is a section from $T\mathcal M$. Force is an important example
whose differential geometric representation is a 1-form; a force 1-form (or
co-vector) acts on a velocity vector to produce power.  Electric field is
another example which can be represented as a 1-form.  Conventionally, in
mechanics, the distinction between a 1-form and a vector is ignored allowing one
to think of both force and velocity as vectors. This lack of distinction between
vectors and 1-forms sacrifices important algebraic and analytical properties
pertaining to the exterior algebra, exterior derivatives and integrability of
differential forms. In particle mechanics, the integrability of 1-forms (exact
forms) translates to the existence of a potential for a force.  An alternative
and geometric approach to understand $n$-form is to think of them as objects
that can be integrated over an orientable $n$ sub-manifold to produce a real
number.  In other words, an $n$-form is a map from an $n-$dimensional
sub-manifold to real numbers. The degree of a differential form is an important
property; it is the dimension of the sub-manifold on which the form has to be
integrated to produce a real number; it can vary from zero to the dimension of
$\mathcal{M}$. The set of all differential forms of degree $n$ over $\mathcal M$
is denoted by $\Lambda^n(\mathcal M)$; we may sometimes suppress the argument
whenever it is clear where the differential forms are defined. The collection of
all $\Lambda^n$ over $n$ is denoted by $\Lambda(\mathcal
M)=\bigcup_{i=1,...,n}\Lambda^i(\mathcal M)$. Here $n$ is the dimension of the
manifold $\mathcal M$. $\Lambda(\mathcal M)$ is often called the exterior
algebra over $\mathcal M$.

An important algebraic operation on differential forms is the wedge or skew
product. Using the wedge product, one can combine two differential forms to
produce a differential form of higher degree.  This algebraic operation is
denoted by $\wedge$; it is antisymmetric and bilinear. It takes two differential
forms of degrees $m$ and $n$ to produce a differential form of degree $m+n$. For
any two differential forms $\alpha$ and $\beta$ of degree $m$ and $n$, the wedge
product between them satisfies the following anti-symmetric relationship.
\begin{equation}
    \alpha\wedge\beta=(-1)^{mn}\beta\wedge\alpha
    \label{eq:wedgeProduct}
\end{equation}
Let $\mathcal{M}$ and $\mathcal{N}$ be two smooth manifolds and
$\varphi$ be a diffeomorphism between them; $\varphi:
\mathcal{M}\rightarrow\mathcal{N}$ and 
$\alpha \in\Lambda^n (\mathcal{N})$. The pull back of
$\alpha$ to the co-tangent space $T^*\mathcal{M}$ is denoted by
$\varphi^*(\alpha)$ and is given by the following relationship,
\begin{equation}
    \varphi^*(\alpha)(v^1,...,v^n)=\alpha(\ed \varphi v^1,...,\ed
    \varphi v^n),\quad v^i\in T\mathcal{M}
    \label{eq:pullBack}
\end{equation}
Here, $\ed \varphi$ is the differential of $\varphi$; it maps $T\mathcal{M}$ to
$T\mathcal{N}$. \eqref{eq:pullBack} must be understood as a point-wise
relationship at each tangent space of $\mathcal M$ and $\varphi(\mathcal M)$
respectively. Under the pull-back map, the wedge product is distributive. 
If $\alpha,\beta \in T^*\mathcal{N}$, then
\begin{equation}
    \varphi^*(\alpha\wedge\beta)=\varphi^*(\alpha)\wedge\varphi^*{\beta}
    \label{eq:pullWedge}
\end{equation}
where $\varphi^*(.)$ is defined as in \eqref{eq:pullBack}. 

On a smooth manifold, one can define a notion of differentiation for differential
forms called the exterior differentiation. It turns out that this notion of
differentiation is coordinate independent. If $\alpha$ is a differential form,
then its exterior derivative is denoted by $\ed \alpha$.  The exterior
derivative operator is a linear map and increases the degree of a differential
form by 1. An important property of the exterior derivative is that $\ed \circ
\ed=0$; the second exterior derivative of a differential form is identically
zero. Under a diffeomorphism, the exterior derivative commutes with its pull back.
This relationship may be written as,
\begin{equation}
    \varphi^*{\ed \alpha}=\ed (\varphi^* \alpha)
    \label{eq:pullExtDer}
\end{equation}
Exterior derivative is also distributive when applied to the wedge product of
two differential forms. For differential forms $\alpha$ and $\beta$ of degree $n$
and $m$, the exterior derivative of wedge product between them is given by,
\begin{equation}
    \ed(\alpha \wedge \beta)=\ed \alpha \wedge \beta + (-1)^n\alpha \wedge \ed \beta
\end{equation}
Using exterior differentiation, one can define closed and exact forms. We say that
a differential form $\alpha\in \Lambda^n$ is exact if there exists a differential
form $\beta\in \Lambda^{n-1}$ such that $\alpha=\ed \beta$. The differential form
$\alpha$ is closed if $\ed \alpha =0$. If we assume $\mathcal M$ to be a
simply connected subset of $\mathbb{R}^n$, Poincar\'e lemma establishes that
closed forms are also exact. The failure of a closed form to be exact is 
measured by the co-homology group. If $\mathcal{C}\subset\mathcal{M}$ is a
hyper-surface of dimension $m$ and $\alpha$ is a differential form of degree
$m-1$, Stokes theorem relates $\ed \alpha$ to the trace of $\alpha$ on
the boundary of $\mathcal{C}$ denoted by $\partial \mathcal{C}$. Stokes theorem
can be written as,
\begin{equation}
    \int_{\mathcal{C}}\ed \alpha=\int_{\partial \mathcal{C}} \alpha
\end{equation}
We now assume that the manifold $\mathcal M$ is equipped with a metric $g$ and
$\ed \mathsf{V}$ denotes the volume form generated by the metric. In
such a case, one can define a linear isomorphism between differential forms of
degree $m$ and $n-m$ ($n$ denotes the dimension of $\mathcal M$), for each $n$.
This isomorphism is defined by the following relationship,
\begin{equation}
    \alpha\wedge\beta=\langle {}_{\star}\alpha,\beta \rangle_{g} \ed \mathsf V;\quad \alpha\in
    \Lambda^n,\beta \in \Lambda^{n-m}
\end{equation}
In the above equation, $\langle.,.\rangle_{g}$ is the inner-product induced by
the metric on $\Lambda^{n-m}$.  This isomorphism is called the Hodge star map;
$\star:\Lambda^m\rightarrow \Lambda^{n-m}$. It turns out that this map is
useful in defining stresses and computing the variation of the HW energy
functional.

%% file: kinematics
\section{Kinematics}\label{sec:kinematics}
The reference configuration of the body is identified with a smooth manifold
with boundary.  This manifold is denoted by $\mathcal{B}$ and its boundary by
$\partial \mathcal{B}$. Similarly, the deformed configuration and its boundary
are denoted by $\mathcal{S}$ and $\mathcal{\partial S}$ respectively. These
configurations are endowed with a $C^\infty$ chart from which they inherit their
smoothness. Following the usual notation in continuum mechanics, we label the
material points of the reference and deformed configurations by their position
vectors. The position vectors of a material point in the reference and deformed
configuration are denoted by $X$ and $x$ with coordinates $X^i$ and $x^i$. The
tangent and co-tangent spaces at each point $X\in \mathcal{B}$ is denoted by
$T_X\mathcal{B}$ and $T^*_X\mathcal{B}$ respectively. The deformation map
relating the reference and deformed configurations is denoted by $\varphi:
\mathcal{B}\rightarrow\mathcal{S}$.  The tangent and co-tangent spaces of the
deformed configuration at a point $x=\varphi(X)$ is denote by
$T_{\varphi(X)}\mathcal{S}$ and $T^*_{\varphi(X)}\mathcal{S}$ respectively.

\subsection{Frame and co-frame fields}
At each tangent space of $\mathcal{B}$, we choose a collection of orthogonal
vectors; we denote these vectors by $E_i$. Note that the orthogonality here is
with respect to the Euclidean inner product. In other words, we have assumed
that each tangent space of the reference or deformed configuration is endowed
with a metric tensor, which is Euclidean. We call this collection of vector
fields a frame field to the reference configuration and it is denoted by
$\mathcal{F}_\mathcal{B}=\{E_1,...,E_n\}$. It is clear that the frame fields at
point $X$ span $T_X\mathcal{B}$. Similarly, the orthonormal frame field
associated with the deformed configuration is denoted by
$\mathcal{F}_\mathcal{S}=\{e_1,...,e_n\}$, where $e_i$ are sections from
$T\mathcal{S}$. The natural (algebraic) duality between tangent and cotangent
spaces induces co-frames on the cotangent bundles of $\mathcal{B}$ and
$\mathcal{S}$. The co-frames of the reference and deformed configurations are
denoted by $\mathcal{F}^*_\mathcal{B}=\{E^i,...,E^n\}$ and
$\mathcal{F}^*_\mathcal{S}=\{e^1,...,e^n\}$ respectively, where $E^i$ and $e^i$
are sections from the cotangent bundles of the reference and deformed
configurations. The natural duality between frame and co-frame fields of the
reference and deformed configurations may be written as,
\begin{equation}
    E^i(E_j)=\delta^i_j;\quad
    e^i(e_j)=\delta^i_j \quad
    E^i\in T^*\mathcal{B}, e^i\in T^*\mathcal{S}
\end{equation}
The differential of the position vector of a material point in the reference
configuration is denoted by $\ed X$ and is given by,
\begin{equation}
    \ed X= E_i\otimes E^i
    \label{eq:refDiff}
\end{equation}
Similarly, the differential of a position vector in the deformed configuration 
in terms of the frame and co-frame fields is
given by,
\begin{equation}
    \ed x= e_i\otimes e^i
    \label{eq:defDiff}
\end{equation}
From the definition of $\ed X$ in \eqref{eq:refDiff}, it can be seen that
tangent vectors from the reference configuration are  mapped to itself under
$\ed X$. To see this, choose $V\in T_X \mathcal{B}$  with $V=c^i E_i$.
Substituting the latter and using the definition of $\ed X$, we arrive at  $\ed
X(V)=c^j E_i E^i(E_j)$. Using the duality between the frame and co-frame fields, we
conclude $\ed X(V)=V$.  Similarly, $\ed x$ maps a tangent vector from the
deformed configuration on to itself.  The differential of a position vector in the
reference and deformed configurations are thus identity maps on the
respective tangent spaces. The importance of the differential of position will be
clarified when we discuss deformation gradient.

Similar to the differential to position, one
can also define the differential of a frame field. The differential of the frame fields 
in the reference configuration is given by,
\begin{equation}
    \ed E_i=\gamma^j_i\otimes E_j
\end{equation}
where, $\gamma^i_j$ is called the connection matrix; it contains 1-forms as its
entries. Because of the orthogonality of the frame fields, the connection
matrix is skew symmetric, i.e. $\gamma^i_j=-\gamma^j_i$. Similarly, the differential of
the frame fields associated with the deformed configuration is given by,
\begin{equation}
    \ed e_i=\bar{\omega}^j_i\otimes e_j
\end{equation}
$\bar{\omega}^i_j$ is the connection matrix associated with the deformed frame
fields and it is also skew symmetric, $\bar{\omega}^i_j=-\bar{\omega}^j_i$.

For a given choice of connection 1-forms and position 1-forms, there are certain
compatibility conditions (Poincar\'e relations) guaranteeing the existence of position
vectors.  These equations are called Cartan's structure equations. The first
compatibility condition establishes the torsion free nature of a configuration.  For
the reference and deformed configurations, this condition may be written as,
\begin{equation}
    \ed^2{X}=0;\;\ed^2 x=0
\end{equation}
Plugging \eqref{eq:refDiff} and \eqref{eq:defDiff} into the above equation leads to,
\begin{equation}
    \ed E^i=\gamma^i_j \wedge E^j;\;
    \ed e^i=\bar{\omega}^i_j \wedge e^j
\end{equation}
The second compatibility condition establishes that the reference and deformed
configurations are curvature-free. This leads to the following conditions on the
reference and deformed frame fields,
\begin{equation}
    \ed^2 E_i =0;\;\; \ed^2e_i=0
\end{equation}
Using the differential of the frame fields in the above equations leads to,
\begin{equation}
    \ed \gamma^i_j= \gamma^i_k\wedge\gamma^k_j;\;\;
    \ed \bar{\omega}^i_j= \bar{\omega}^i_k\wedge\bar{\omega}^k_j
\end{equation}
For a simply connected body, the structure equations as above for the reference
and deformed configurations provide the necessary kinematic closure to ensure
that the configurations can be embedded with an Euclidean space. Indeed, without
this closure effected by the structure equations, a model cannot in general
produce a deformed configuration which is Euclidean embeddable.

\subsection{Differentials of position and frame}
\begin{figure}
    \centering
    \includegraphics[scale=.95]{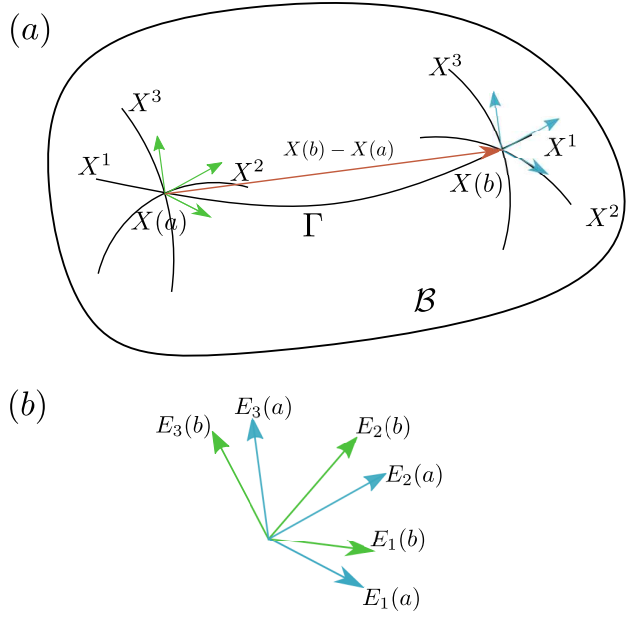}
    \caption{The coordinate lines and the frame field generated from these
    coordinate lines are shown in (a). The frame fields at $X(a)$ and
    $X(b)$ are shown in (b); we have moved the frames to the same point so that
    it is convenient to interpret the change. We have used the
    notation $E_i(a)$ and $E_i(b)$ to indicate the frame at the material
    points $X(a)$ and $X(b)$.}
    \label{Fig:cordinateSystem}
\end{figure}

In the previous sub-section, we introduced the notion of differentials to
position vectors and frame fields. We now present the geometric meaning of these
infinitesimal quantities.  Consider the differential of the position vector in
the reference configuration given in \eqref{eq:refDiff}. For a given coordinate
system, the position vector $X$ is a smooth function of its coordinates
$(X^1,X^2,X^3)$.  Let $\Gamma$ be a parametrized curve, $\Gamma:[a,b]\rightarrow
\mathcal{B}$. For convenience, we assume $\Gamma$ to be along the first
coordinate direction, which is obtained by freezing the other two coordinate
functions to some constant.  We also assume that the frame fields $E_i$ are
constructed by a Gram–Schmidt procedure on the tangent vectors of the
co-ordinate lines at each material point $X$. Fig. \ref{Fig:cordinateSystem}
shows how the position vector and the frame change as one moves along the curve
$\Gamma$. From our assumptions on $\Gamma$, it may be seen
that the tangent vector to  $\Gamma$ is given by the $cE_1$, where $c$ is a real
valued function.  Now, the differential of position, which is a vector valued
1-form, can be integrated along the curve $\Gamma$ to produce a vector. This is
nothing but the vector between the material points $X(a)$ and
$X(b)$, which may be formally written as,
\begin{align}
    \nonumber
    X(b)-X(a) & =\int_a^b \ft{d}X(cE_1)\ed s     \\
    \nonumber
              & =\int_a^b cE_i E^i(E_1)\ed s,    \\
              & =\int_a^b cE_1\ed s,
\end{align}
In the above equation, $E_1$ and $c$ can vary along the curve $\Gamma$. The
above interpretation of $\ed X$ is very similar to that of 1-forms as real
numbers defined on curves.

We now consider the differential of the frame field, integrating which along
$\Gamma$ leads to the relative rotation of the frame $\mathcal{F}$ between the
points $X(a)$ and $X(b)$. Fig. \ref{Fig:cordinateSystem}(b) demonstrates the
rotation experienced by the frame as one traverses along the curve $\Gamma$. 
\begin{align}
    \nonumber
    E_i&=\int_a^b\ed E_i(E_1)\ed s;\\
    &=\int_a^bE_j\gamma^j_i(E_1)\ed s;
\end{align}
From the discussion, it is clear that the displacement (rotation) vector
between two points (frames) in a configuration depends on the position 1-form and the
curve or path used for integration. However, the displacement vector
between two points on a configuration should be independent of the path chosen to
integrate the 1-forms.  This condition is exactly what the structure
equation enforces.

\subsection{Deformation gradient}
As discussed earlier, the deformation  map  sends the position vector of a
material point in the reference configuration to its corresponding position
vector in the deformed configuration.  The differential of the deformation map
or the deformation gradient, denoted by $\ed \varphi$, maps the tangent space of
the reference configuration to the corresponding tangent space in the deformed
configuration.  For an assumed frame field (for both reference and deformed
configurations), the differential of the deformation map can be obtained by
pulling back the co-vector part of the differential of the deformed position
vector.
\begin{align}
    \nonumber
    \ft{d}\varphi & =e_i\otimes \varphi^*(e^i)                                                       \\
                  & =e_i\otimes\theta^i; \quad \theta^i \in T^*\mathcal{B},e_i\in\mathcal{F}_\mathcal{S}
    \label{eq:deformationOneForm}
\end{align}
In writing \eqref{eq:deformationOneForm}, we have introduced the following
definition: $\theta^i:=\varphi^*({e^i})$.  In our construction, the 1-forms
$\theta^i$ contain local information about the deformation map $\varphi$. A
primitive variable in our theory, we call this the deformation 1-form.  From
\eqref{eq:deformationOneForm}, we see that the vector leg of the deformation
gradient is from the deformed configuration, while the co-vector leg is from the
reference configuration. If $V(X)\in T_X\mathcal{B}$, the action of the
deformation gradient on $V$ is given by,
\begin{equation}
    \ed\varphi(V)=e_i\theta^i(V)
\end{equation}
Since $\theta^i(V)$ are real numbers, the above equation is a linear combination of
tangent vectors from the deformed configuration.

\subsection{Pull-back of structure equations}
Having introduced the deformation gradient and the deformation 1-form in the
last subsection, we are now ready to rewrite the referential version of
structure equations in the deformed configuration. These equations are obtained
by pulling back the structure equations for the deformed configuration under the
deformation map:
\begin{align}
    \nonumber
    \varphi^*(\ed{e^i} )& = \varphi^*(\bar{\omega}^i_j\wedge e^j) \\
    \ed \theta^i           & =\omega^i_j\wedge \theta^j
\end{align}
where, $\omega^i_j:=\varphi^*(\bar{\omega}^i_j)$ are called pulled back
connection 1-forms. The key fact used in obtaining the above equation is that
exterior derivative and wedge product commute with pull back. Using a similar
argument, the pull back of the second compatibility condition is
given by,
\begin{equation}
    \ed \omega^i_j=\omega^i_k\wedge\omega^k_j
\end{equation}
For concreteness, we present the components of the structure equations in three
spatial dimensions. Each structure equation constitutes a system of three
equations.  The matrix form of the first structure equation (referentially
pulled back) may be written as,
\begin{equation}
    \begin{bmatrix}
        \ft{d}\theta^1 \\
        \ft{d}\theta^2 \\
        \ft{d}\theta^3 \\
    \end{bmatrix}
    =
    \begin{bmatrix}
        0           & \omega^1_2 & -\omega^1_3 \\
        -\omega^1_2 & 0          & -\omega^2_3 \\
        \omega^1_3  & \omega^2_3 & 0
    \end{bmatrix}
    \wedge
    \begin{bmatrix}
        \theta^1 \\
        \theta^2 \\
        \theta^3 \\
    \end{bmatrix}
    \label{eq:FirstCompatablityMatrix}
\end{equation}
Note that the  combining rule for the elements of the matrix and the vector on
the right hand side of the above equation is via the wedge product.
The component form of the second compatibility equations is given by,
\begin{align}
    \nonumber
    \ed\omega^1_3 & =\omega^1_2\wedge\omega^2_3 \\
    \ed\omega^2_3 & =\omega^1_2\wedge\omega^1_3 \\
    \nonumber
    \ed\omega^1_2 & =\omega^1_3\wedge\omega^2_3
\end{align}

\subsection{Strain and deformation measures}
The notion of length is central to continuum mechanics; important kinematic
quantities like strain and rate of deformation are derived from it. Indeed, it
may not be possible to assess the state of deformation without the metric
structure defined on both reference and deformed configurations. The metric
structure of a configuration is defined by a symmetric and positive definite
tensor, which encodes the notion of length (in that configuration).  We denote
the metric tensor of the reference and deformed configurations by $G$ and $g$
respectively;  $G:T_X\mathcal{B}\times T_X\mathcal{B}\rightarrow\mathbb{R}$ and
$g:T_x\mathcal{S}\times T_x\mathcal{S}\rightarrow\mathbb{R}$. These tensor
valued functions pertain to the idea of infinitesimal lengths at the tangent
spaces of $\mathcal{B}$ and $\mathcal{S}$. As such, the notion of length is an
additional structure placed on $\mathcal{B}$ and $\mathcal{S}$. In this work, we
assume these metrics to be Euclidean. In terms of the co-frame field, the metric
tensor of the reference configuration is given as,
\begin{align}
    \nonumber
    G & =(E^j\otimes E_j):(E_i\otimes E^i)\\
      & =E^i\otimes E^i
\end{align}
The double contraction in the above equation is calculated using the inner
product induced by the Euclidean metric.  Similarly, the metric
tensor in the deformed configurations may be written as,
\begin{align}
    \nonumber
    g & =(e^i\otimes e_j):(e_j\otimes e^i) \\
      & =e^i\otimes e^i
\end{align}
In terms of the frame fields, the inverses of the metric tensors for the
reference and deformed configurations may be written as,
\begin{equation}
    G^{-1}=E_i\otimes E_i;\;\;\;g^{-1}=e_i\otimes e_i
\end{equation}
The right Cauchy-Green deformation tensor is obtained via the pull back of the
metric tensor of the deformed configuration to the reference configuration. In
terms of the the co-frame fields, this relationship can be written as,
\begin{align}
    \nonumber
    C & =\varphi^*(g)                       \\
    \nonumber
      & =\varphi^*(e^i\otimes e^i) \\
      & =\theta^i\otimes\theta^i
    \label{eq:rightCGMetric}
\end{align}
An alternate way to compute the $C$ is to use the usual definition
in continuum mechanics,
$C=\ft{d}\varphi^t\ft{d}\varphi$. Here, the $(.)^t$ is understood to be the
adjoint map induced by the metric structure. Using the orthonormality
of the frame field we arrive at,
\begin{align}
    \nonumber
    C & =(\theta^i\otimes e^i) (e_j\otimes \theta^j) \\
      & =\theta^i\otimes\theta^i
    \label{eq:rightCGDefGrad}
\end{align}
The calculations leading to \eqref{eq:rightCGMetric} and \eqref{eq:rightCGDefGrad} are
exactly the same; only the sequence in which pull back and inner product are applied
differs.  The Green-Lagrangian strain tensor may now be written as,
\begin{align}
    \nonumber
    E & =\frac{1}{2}(C-G)                                                     \\
      & =\frac{1}{2}[(\theta^i\otimes\theta^i)-(E^i\otimes E^i)]
\end{align}
The the first invariant of the right Cauchy-Green tensor is given by,
\begin{equation}
    I_1=\langle \theta^i,\theta^i \rangle_{G}
\end{equation}
Here $\langle.,.\rangle_G$ denotes the inner product induced
by the metric tensor $G$. The area forms induced by the co-frame of the
reference configuration are given by,
\begin{equation}
    \ed A^1=E^2\wedge E^3;\quad
    \ed A^2=E^3\wedge E^1;\quad
    \ed A^3=E^1\wedge E^2;\quad
\end{equation}
Similarly, the area forms induced by the co-frame of the deformed configurations are
given by,
\begin{equation}
    \ed a^1=e^2\wedge e^3;\quad
    \ed a^2=e^3\wedge e^1;\quad
    \ed a^3=e^1\wedge e^2;\quad
\end{equation}
We also define the pulled back area forms from those of the
deformed configuration to the reference configuration. These area forms are
obtained as,
\begin{equation}
    \ed \mathsf{A}^1=\theta^2\wedge \theta^3;\quad
    \ed \mathsf{A}^2=\theta^3\wedge \theta^1;\quad
    \ed \mathsf{A}^3=\theta^1\wedge \theta^2;\quad
    \label{eq:pullBackAreaForm}
\end{equation}
In writing the above equations, we have used $\ed \mathsf{A}^i:=\varphi^*(\ed
a^i)$ and the definition of the deformation 1-form. The second invariant of $C$
is now given by,
\begin{equation}
    I_2=\langle\ed \mathsf{A}^i,\ed \mathsf{A}^i \rangle_G
\end{equation}
In terms of co-frame fields, the volume forms of reference and deformed
configurations may be written as,
\begin{equation}
    \ed V=E^1\wedge E^2 \wedge E^3;\quad
    \ed v=e^1\wedge e^2 \wedge e^3
\end{equation}
The pull back of $\ed v$ to the reference configuration is,
\begin{equation}
    \ed \mathsf{V}=\theta^1\wedge\theta^2\wedge \theta^3
\end{equation}
The third invariant of $C$ is then given by,
\begin{equation}
    I_3=({}_{\star}\ed \mathsf{V})^2
\end{equation}

\subsection{Affine connection via frame fields}
An affine connection on a smooth manifold is a device used to differentiate
sections of vector and tensor bundles in a co-ordinate independent manner. This
differentiation is often referred to as covariant.  The choice of an
affine connection determines the covariant differentiation. For a smooth
manifold with a metric, the metric induces a unique covariant derivative and we
denote it by $\nabla$. For vector fields $v, w$ and a real valued function $f$, the
covariant derivative satisfies the following properties.
\begin{align}
    \nonumber
    \nabla_{fe_i}w    & =f\nabla_{e_i}w;\;\; f\in \Lambda^0 \\
    \label{eq:covProp}
    \nabla_{e_i}(w+v) & =\nabla_{e_i}w+\nabla_{e_i}v        \\
    \nonumber
    \nabla_{e_i}(fw)  & =e_i[f]w+f\nabla_{e_i}w
\end{align}
$e_i[f]$ in the above equations denotes the action of a vector field on a real
valued function $f$.  We now show how the connection 1-forms encode the affine
connection.  Let $w=\sum_{i=1}^2w^ie_i$ is an arbitrary vector
field defined on $\mathcal{S}$. Then the covariant derivative of $w$ in the direction of
$e_i$ is given by,
\begin{equation}
    \nabla_{e_i} w= \ed w^j(e_i)e_j+w^j\bar{\omega}^k_j(e_i)e_k
\end{equation}
Having defined the covariant derivative of a vector field in the direction of
frame fields, it is now possible to extend the above definition of covariant
differentiation to arbitrary vector fields using the defining properties given
in \eqref{eq:covProp}.

%% file: stress
\section{Stress as co-vector valued two-form} \label{sec:stress} 
In the previous section, we have reformulated the kinematics of continua using
differential forms. The right Cauchy-Green deformation tensor and the
deformation gradient were the key objects in the reformulation. In this section,
we present a geometric approach to stress, originally due to Kanso \textit{et
al.} \cite{kanso2007}. Even though this approach is intuitive and geometric, it
was never put to use.  In classical dynamics \cite{marsden2013}, force is
understood as a co-vector so that its pairing with velocity produces power.
This understanding of force as dual to velocity does not require the metric
tensor to compute power and it must be contrasted with the usual understanding
of of both velocity and force as vectors. Extending this concept of force as a
co-vector to the continuum mechanical definition of stress is our present goal.
As a consequence, we also show that the description of power/work used in
continuum mechanics may be undertaken without the notion of a metric.  However,
it should also be noted that a formulation of continuum mechanics without the
metric tensor may not be feasible as the notion of strain crucially depends on
the metric.

We denote the Cauchy stress tensor by $\sigma$. The traction 'vector' acting on
an infinitesimal area with unit normal $n$ is denoted by $t$, given by the well
known formula $t=\sigma n$. The traction $t$ is a force which depends on the
material point at which it is evaluated and the area which sustains it. The
relation between the normal and the traction is postulated to be linear. In the
language of differential forms, an infinitesimal area is regarded as a 2-form,
while from classical dynamics we also know that force is a co-vector or a
1-form. Putting these two ideas together, we are led to a geometric definition
of Cauchy stress given by,
\begin{equation}
    \sigma=t^i\otimes \ft{d}a^i;
    \label{eq:Cauchy3Tensor}
\end{equation}
In the above equation, $\ed a^i$ is the area 2-form, which sustains the traction
vector $t^i$. The tensor product in the above equation is due to the linearity
between the traction and area forms. From this equation, it is easy to see that
the area forms change orientation if the order the co-vectors are reversed.
Geometrically, if there are $n$ linearly independent area forms at a point on
the manifold, the stress tensor assigns to each area form a 1-form called
traction. With this understanding, Cauchy stress may now be identified with a
section from the tensor bundle $\Lambda^1\otimes\Lambda^2$ which has the
deformed configuration as its base space.

In contemporary continuum mechanics, Nanson's formula describes how unit normals
transform under the deformation map \cite{ogden1997}. Geometrically, Nanson's
formula is nothing but the pull back of a 2-form under the deformation map.  The
pull back of area forms generated by the co-frame is already described in
\eqref{eq:pullBackAreaForm}.
\begin{figure}
    \centering
    \includegraphics[scale=0.5]{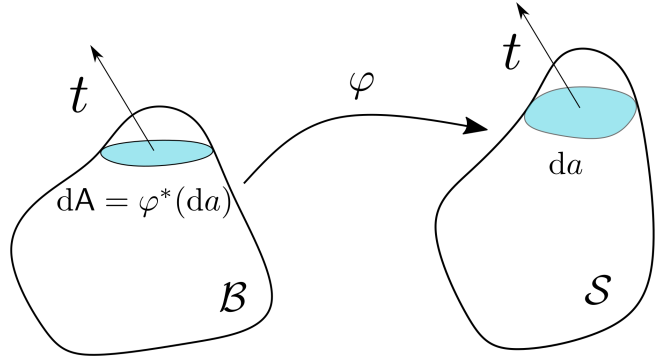}
    \label{Fig:piolaTransform}
    \caption{Piola Transform: notice that the length and direction of the force acting an the
    deformed and reference area is preserved under Piola transform.}
\end{figure}
The first Piola stress may now be obtained by pulling back the area leg of
the Cauchy stress under the deformation map. This partial pull back of the Cauchy
stress is termed the Piola transform. This
relationship may be formally written as,
\begin{align}
    \nonumber
    P & =t^i\otimes \varphi^*(\ft{d}a^i) \\
      & =t^i\otimes \ed \mathsf A^i
    \label{eq:Piola3tensor}
\end{align}
Note that the traction 1-form in both the definitions of Cauchy stress and first
Piola stress are the same, i.e. the pull back does nothing to the traction
1-form.  In the above discussion, contrary to convention, Cauchy and first Piola
stresses are identified as third order tensors, not the usual second order. This
ambiguity can be removed if one applies Hodge star on the area leg of these two
stresses.  In three dimensions, the Hodge star establishes an isomorphism
between forms of degree 2 and 1. Applying Hodge star to Cauchy and first Piola
stresses leads to,
\begin{align}
    \label{eq:Cauchy2Tensor}
    \sigma & =t^i\otimes ({}_{\star}\ft{d}a)^i \\
    \label{eq:Piola2Tensor}
    P      & =t^i\otimes ({}_{\star}\ft{d}\mathsf A)^i
\end{align}
Kanso \textit{et al.} made a distinction between the stress tensors given in
\eqref{eq:Cauchy3Tensor}, \eqref{eq:Piola3tensor} and \eqref{eq:Cauchy2Tensor},
\eqref{eq:Piola2Tensor}. However we do not see a need for it, since both the
usual and geometric definitions of stress contain exactly the same
information; only the ranks of these tensors are different.

\subsection{Traction 1-form via a stored energy function}
The Doyle-Ericksen formula is an important result in continuum
mechanics \cite{doyle1956nonlinear}, which states that the Cauchy stress
understood as a 2-tensor and the metric tensor of the deformed configuration
are work conjugate pairs. For a stored energy density function $W$, Doyle-Ericksen
formula gives us the following relationship,
\begin{equation}
    \sigma=2\frac{\partial W}{\partial g}
    \label{eq:doyleErichsen}
\end{equation} 
In writing \eqref{eq:doyleErichsen}, we have assumed that $W$ depends on the
differential of deformation through the right Cauchy-Green deformation tensor,
i.e. $W$ is frame-indifferent. From the discussion presented in the previous
section, it can be seen that the area leg of the Cauchy stress tensor is
determined by the choice of coordinate system for the tangent bundle of the
deformed configuration.  On the other hand, the area leg of the first Piola
stress is determined by both the co-ordinate system for the tangent bundle of
the deformed configuration and the deformation map. Clearly, the area leg of a
stress tensor does not require a constitutive rule, the only component requiring
so is the traction 1-form.

We now claim that for a stored energy function $W$, the traction 
$1-$form has the following constitutive rule, 
\begin{equation}
    \label{eq:ConstitutionTraction}
    t^i=\frac{1}{J}\frac{\partial W}{\partial e_i}
\end{equation}
The last equation is in the same spirit as the Doyle-Ericksen formula. To
establish the result stated in \eqref{eq:ConstitutionTraction}, we first compute
the directional derivative of $W$ along $e_i$,
\begin{subequations}
\label{eq:tiConjei}
\begin{align}
    \label{eq:chainRule}
    \frac{\partial W}{\partial e_i}&=\frac{\partial W}{\partial (\ed \varphi)}
    \frac{\partial (\ed \varphi)}{\partial e_i}\\
    \label{eq:piolaTwoTensor}
    &=(t^j\otimes {}_{\star}\ed \mathsf A^k)(e^i\otimes e_j \otimes \theta^k)\\
    \label{eq:tiConjei-5}
    &=\langle {}_{\star}\ed \mathsf{A}^J, \theta^J \rangle t^i\\
    &=t^i {}_{\star}\ed \mathsf{V}\\
    &=Jt^i
\end{align}
\end{subequations}
We used chain rule to arrive at the right hand side of \eqref{eq:chainRule}.  In
\eqref{eq:piolaTwoTensor}, the expression for Piola stress (as a two tensor) in
terms of $W$ and the directional derivative of $\ed \varphi$ along 
$e_i$ are used to get the right hand side and performing the required
contractions lead to \eqref{eq:tiConjei-5}. The claim is finally established by
using the definitions of pull back and Hodge star for volume forms.  In
three dimensions, constitutive relations have to be supplied to the three traction
1-forms each conjugate to vector elements in the frame to close the equations of
motion. If the Cauchy stress generated by a stored energy function is known,
then the expression for traction 1-form can be computed using the simple
relationship $t^i=\sigma n^i$, where the vector fields $n^i$ are chosen to be
elements from the frame of the deformed configuration.

\section{Hu-Washizu variational principle}\label{sec:variational}
We first present the conventional HW variational principle for a finitely
deforming elastic body. The HW variational principle takes the deformation
gradient, first Piola stress and deformation map as input arguments. The
referential version of HW functional for a non-linear elastic solid reads,
\begin{equation}
    \label{eq:conventionalHW}
    I_{HW}=\int_{\mathcal{B}}[W(C)-P:(F-\ed \varphi)]\ed V- \int_{\partial
    \mathcal{B}}\langle t,\varphi \rangle \ed A
\end{equation}
In the above equation $t=PN$, is the traction defined on the surface $\partial
\mathcal{B}$.
The integration in the above equation is withrespect to the volume and the area
forms of the reference configuration. 
In \eqref{eq:conventionalHW}, deformation gradient is assumed to
be independent, this tensor field is denoted by $F$, while the deformation
gradient computed from deformation is denoted by $\ed \varphi$. Another
important point to notice is that the the second term in the above equation is
bilinear in Piola stress and deformation gradient. The variation of the HW
functional with respect to deformation, deformation gradient and first Piola
stress leads to the equilibrium equation, constitutive rule and compatibility of
deformation gradient. HW variational principle have been previously exploited to
formulate numerical solution procedure to solve non-linear problems in
elasticity see\cite{angoshtari2017,shojaei2019,shojaei2018}.

\subsection{Hu-Washizu variational principle using geometric definitions of
stress and deformation}
From now on we will use the definitions of Cauchy and Piola stresses as
given in \eqref{eq:Cauchy3Tensor} and \eqref{eq:Piola3tensor} respectively.
We now rewrite the HW variational principle such that it takes deformation
1-forms, traction 1-forms and deformation map as inputs. We
also assume that compatible frame fields for the reference and deformed
configurations are given. The HW functional may be written as,
\begin{equation}
    I(\theta^i,t^i,\varphi)=\int_{\mathcal{B}}W(\theta^i)\ed V-(t^i\otimes
    \ed \mathsf{A}^i)\dot{\wedge}(e_i \otimes (\theta^i- \ft{d}\varphi^i))
    -\int_{\partial \mathcal{B}}\langle t^\sharp,\varphi \rangle\ed \mathsf{A}
    \label{eq:modifiedHW0}
\end{equation}
Note that we did not write the volume form in the second term on the RHS of
\eqref{eq:modifiedHW0}, since the outcome of $\dot{\wedge}$ is a 3-form which
can be integrated over the reference configuration to produce work done by the
traction 1-forms on the frame fields. In \eqref{eq:modifiedHW0},
$\dot{\wedge}$ denotes a bilinear map. For $\alpha \in T^*\mathcal{S}$, $v\in
T\mathcal{S}$ and $a,b \in \Lambda(\mathcal{B})$ the action of this map is given
by $(\alpha\otimes a)\dot{\wedge}(v\otimes b)=\alpha(v)a\wedge b$. Note that the
definition of $\dot{\wedge}$ given here is a little different from the
one in Kanso \textit{et al.}; specifically, we do not use the metric tensor. From
the definition of $\dot{\wedge}$, it can be seen that the work done by
stress on deformation is metric independent. This property of our current
variational formulation
brings the continuum mechanical definition of stress a step closer to the
definition of force as defined in classical mechanics. Another important point
here is that the second term on the RHS in \eqref{eq:modifiedHW0} is equivalent to
the second term in \eqref{eq:conventionalHW}; however now the relationship
between the different arguments is multi-linear.

\textit{Remark 1:} In writing \eqref{eq:modifiedHW0}, we have presumed that the
geometry of the body is Euclidean and it is frozen during the deformation
process. We believe that this assumption can be relaxed by permitting
non-integrability in the connections and deformation 1-forms (i.e. by incorporating source
terms in the structure equations).

\textit{Remark 2:} For a frame to represent Euclidean geometry, it is not required
that the connection 1-forms be identically zero. It only requires zero
source term in Cartan's structure equations. Any set of (unit) tangent vectors to a
curvilinear co-ordinate system will have non-zero connection 1-forms, even as it
acts as a frame in the Euclidean space.

We now proceed to obtain the Euler-Lagrange equations or the condition for
critical points of the functional $I$. We use the Gateaux derivative for this
purpose. Let $\epsilon$ denote a small parameter and $\hat{(.)}$ an
increment in quantity with which we are differentiating the functional $I$.
We first
calculate the variation of $I$ with respect to traction 1-forms;
$t^i\mapsto t^i+\epsilon \hat{t}^i$, where $\hat{t}^i$ are assumed
to be from the tangent space of $T^*\mathcal{S}$.
\begin{equation}
    I(\epsilon)=\int_{\mathcal{B}}W(\theta^i)-((t^i+\epsilon \hat{t}^i)\otimes
    \ed \mathsf{A}^i)\dot{\wedge}(e_j \otimes (\theta^j- \ft{d}\varphi^j))
    -\int_{\partial \mathcal{B}}\langle t^\sharp,\varphi \rangle\ed \mathsf{A}
\end{equation}
Using the definition of $\dot{\wedge}$ and Gateaux derivative, we get a vector
valued 3-form for each $i$. These three 3-forms have to be equated to zero to
get the condition for critical points in the direction of traction 1-forms.
These conditions can be formally written as,
\begin{equation}
   \begin{bmatrix}
       (\ed \mathsf{A}^1 \wedge (\theta^1-\ed \varphi^1))
        &-(\ed \mathsf{A}^1\wedge \ed \varphi^2)
        &-(\ed \mathsf{A}^1\wedge\ed \varphi^2)\\
        -(\ed \mathsf{A}^2 \wedge \ed \varphi^1)
        &(\ed \mathsf{A}^2\wedge (\theta^2-\ed \varphi^2))
        &-(\ed \mathsf{A}^2\wedge\ed \varphi^2)\\
        -(\ed \mathsf{A}^3 \wedge \ed \varphi^1)
        &-(\ed \mathsf{A}^3\wedge \ed \varphi^2)
        &(\ed \mathsf{A}^3\wedge(\theta^3-\ed \varphi^3))
    \end{bmatrix}
    \otimes
    \begin{bmatrix}
        e_1\\
        e_2\\
        e_3
    \end{bmatrix}
    =
    \begin{bmatrix}
        0\\
        0\\
        0
    \end{bmatrix}
\end{equation}
Since $e_i$ are elements form the frame, we have $g(e_i,e_i)=1$. The above
equation can be true only when the coefficient matrix on the LHS is zero, which
leads to,
\begin{subequations}
 \begin{align}
     (\ed \mathsf{A}^1 \wedge (\theta^1-\ed \varphi^1))=0;\quad
        (\ed \mathsf{A}^1\wedge \ed \varphi^2)=0;\quad
        (\ed \mathsf{A}^1\wedge\ed \varphi^2)=0\\
        (\ed \mathsf{A}^2 \wedge \ed \varphi^1)=0;\quad
        (\ed \mathsf{A}^2\wedge (\theta^2-\ed \varphi^2))=0;\quad
        (\ed \mathsf{A}^2\wedge\ed \varphi^2)=0\\
        (\ed \mathsf{A}^3 \wedge \ed \varphi^1)=0;\quad
        (\ed \mathsf{A}^3\wedge \ed \varphi^2)=0;\quad
        (\ed \mathsf{A}^3\wedge(\theta^3-\ed \varphi^3))=0
\end{align}
\end{subequations}
The above equations can be recast in a matrix form as,
\begin{equation}
    \begin{bmatrix}
        \theta^1-\ed \varphi^1 & \ed \varphi^2 & \ed \varphi^3\\
        \theta^1 & \theta^2-\ed \varphi^2 & \ed \varphi^3\\
        \theta^1 & \ed \theta^2 & \ed \theta^3-\varphi^3
    \end{bmatrix}
    \wedge
    \begin{bmatrix}
    \theta^2 \wedge \theta^3\\
    \theta^3 \wedge \theta^1\\
    \theta^1 \wedge \theta^2\\
    \end{bmatrix}
    =
    \begin{bmatrix}
        0\\
        0\\
        0
    \end{bmatrix}
\end{equation}
For these equations to hold, the following
conditions must be met, 
\begin{equation}
    \theta^1-\ed \varphi^1=0;\quad
    \theta^2-\ed \varphi^2=0;\quad
    \theta^3-\ed \varphi^3=0;\quad
    \label{eq:compatabilityEL}
\end{equation} 
The above condition simply states that there exist three zero forms whose
exterior derivatives are the deformation 1-forms; or in other words,
deformation 1-forms are exact and $\varphi^i$ are the potentials for the
corresponding deformation 1-forms.

We now compute the variation of $I$ with respect to deformation 1-forms. 
Incremental changes in the deformation 1-forms may be written as,
$\theta^i\mapsto \theta^i+\epsilon \hat{\theta}^i$, where $\epsilon
\hat{\theta}^i$ is assumed to be an element from the tangent space of
$T^*\mathcal{B}$.
\begin{equation}
    I(\epsilon)=\int_{\mathcal{B}}
    W(\theta^i+\epsilon\hat{\theta}^i)- 
    (t^i\otimes \ed \mathsf{A}(\epsilon)^i)\dot{\wedge}(e_j\otimes \theta^j(\epsilon))
    -\int_{\partial \mathcal{B}}\langle t^\sharp,\varphi \rangle\ed A
\end{equation}
Using the definition of Gateaux derivative, for each $\theta^i$ we have,
\begin{subequations}
    \label{eq:ElConstitutionRaw}
    \begin{align}
        \nonumber
        \frac{\partial W}{\partial \theta^1}
        &=[t^1(e_1){}_{\star}(\theta^2\wedge\theta^3)
        -t^2(e_1){}_{\star}(\ed \varphi^1\wedge\theta^3)
        +t^2(e_2){}_{\star}((\theta^2-\ed \varphi^2)\wedge \theta^3)
        -t^2(e_3){}_{\star}(\ed \varphi^3\wedge\theta^3)\\
        &-t^3(e_1){}_{\star}(\theta^2\wedge\varphi^1)
        +t^3(e_2){}_{\star}(\theta^2\wedge\ed\varphi^3) 
        +t^3(e_3){}_{\star}(\theta^2\wedge(\theta^3-\ed \varphi^3))
        ]^\sharp\\
        \nonumber
        \frac{\partial W}{\partial \theta^2}
        &=[t^1(e_1){}_{\star}(\theta^3\wedge(\theta^1-\ed \varphi^1))
        -t^1(e_2){}_{\star}(\theta^3\wedge\ed \varphi^2)
        -t^1(e_3){}_{\star}(\theta^3\wedge\ed \varphi^3)
        -t^2(e_2){}_{\star}((\theta^3\wedge\ed \theta^1)\\
        &-t^3(e_1){}_{\star}(\ed \varphi^1\wedge\theta^1)
        -t^3(e_2){}_{\star}(\theta^2\wedge \theta^1)
        +t^3(e_3){}_{\star}((\theta^3-\ed \varphi^3)\wedge\theta^1) ]^\sharp\\
        \nonumber
        \frac{\partial W}{\partial \theta^3}
        &=[t^1(e_1){}_{\star}((\theta^1-\ed \varphi^1)\wedge\theta^2)
        -t^1(e_2){}_{\star}(\varphi^2\wedge\ed\theta^2)
        -t^1(e_3){}_{\star}(\varphi^3\wedge\ed \theta^2) 
        -t^2(e_1){}_{\star}(\ed \theta^1\wedge\ed \varphi^1)\\
        &-t^2(e_2){}_{\star}(\theta^1-(\theta^2-\varphi^2))
        +t^2(e_3){}_{\star}(\theta^1\wedge\ed\varphi^3)
        +t^3(e_3){}_{\star}(\theta^1\wedge\theta^2) ]^\sharp
    \end{align}
\end{subequations}
If we now take into account the compatibility equations previously established
in \eqref{eq:compatabilityEL}, the last equations reduce to,
\begin{subequations}
    \label{eq:ElConstitution}
    \begin{align}
        \frac{\partial W}{\partial \theta^1}
        &=[t^1(e_1){}_{\star}(\theta^2\wedge\theta^3)
        +t^2(e_1){}_{\star}( \theta^3\wedge\theta^1)
        +t^3(e_1){}_{\star}(\theta^2\wedge\theta^1)
        ]^\sharp\\
        \frac{\partial W}{\partial \theta^2}
        &=[t^1(e_2){}_{\star}(\theta^2\wedge \theta^3)
        +t^2(e_2){}_{\star}((\theta^3\wedge \theta^1)
        +t^3(e_2){}_{\star}(\theta^1\wedge \theta^2)]^\sharp\\
        \frac{\partial W}{\partial \theta^3}
        &=[t^1(e_3){}_{\star}(\theta^2\wedge\ed \theta^3) 
        +t^2(e_3){}_{\star}(\theta^3\wedge\ed\theta^1)
        +t^3(e_3){}_{\star}(\theta^1\wedge\theta^2) ]^\sharp
    \end{align}    
\end{subequations}
From these equations, it is seen that a 2-form accompanies the components of
traction 1-forms; this is indeed true since we use a Piola transform to write
the constitutive rule in the reference configuration. From a comparison of
\eqref{eq:ElConstitutionRaw}, where Cartan's structure equations for kinematic
closure have not been enforced, and \eqref{eq:ElConstitution}, we note that the
expressions for traction in the former have additional terms. These additional
terms may be related to incompatibilities created by the emergence of defects
(such as dislocations) as the deformation evolves. In other words, without an
explicit imposition of the kinematic closure conditions on the flow, the
deformed body may never be realized as a subset of the Euclidean space.

Finally, we compute the variation of $I$ with respect to deformation;
$\varphi^i\mapsto\varphi^i+\epsilon\hat{\varphi}^i$, where $\hat{\varphi}$
belongs to $T\mathcal{B}$. Using the definition of the superimposed incremental
deformation in $I$ and upon computing the Gateaux derivative, we have the
following equation,
\begin{equation}
    \label{eq:ELEquilibrioumRaw}
    \int_{\mathcal{B}}(t^i\otimes\ed \mathsf{A}^i)\dot{\wedge}(e_j\otimes \ed\hat{\varphi}^i)=0
\end{equation}
To complete the variation, we need to shift the differential from
$\hat{\varphi}$. We first calculate the following,
\begin{equation}
    \label{eq:divTheorem}
    \ed (\varphi^k t^j(e_k)\ed \mathsf{A}^j)=\ed \varphi^k \wedge t^j(e_k)\ed
    \mathsf{A}^j+ \varphi^k\ed (t^j(e_k))\wedge\ed \mathsf{A}^j+\varphi^k
    t^j(e_k)\ed^2\mathsf{A}^j
\end{equation}
This equation invites a few comments. The first is that we are calculating the
exterior derivative of a 2-form, with $\varphi^k t^j(e_k)$ being a scalar. Using
the product rule of differentiation, we have expanded the right hand side of
\eqref{eq:divTheorem}. The second term in \eqref{eq:divTheorem} should be
evaluated using the connection 1-forms since it involves the exterior derivative
of a vector. This terms is relevant when one works with a non-trivial connection
for the manifold, examples being a body with dislocations and a Kirchhoff shell.
If we invoke compatibility of deformation, we have $\ed^2A=0$, which leaves
\eqref{eq:divTheorem} in the following form,
\begin{equation}
    \label{eq:divTheoremWithoutInc}
    \ed (\varphi^k t^j(e_k)\ed \mathsf{A}^j)=\ed \varphi^k \wedge t^j(e_k)\ed
    \mathsf{A}^j+ \varphi^k\ed (t^j(e_k))\wedge\ed \mathsf{A}^j
\end{equation}
An expression similar to \eqref{eq:divTheoremWithoutInc} was utilized by Kanso
\textit{et al.} \cite{kanso2007} to define mechanical equilibrium.  The
expression for exterior derivative defined in \eqref{eq:divTheoremWithoutInc}
involves the connection 1-forms of the manifold, which is similar to the
covariant exterior derivatives used in gauge theories in physics. Using
\eqref{eq:divTheorem} in \eqref{eq:ELEquilibrioumRaw} leads to,  
\begin{equation}
    \int_{\mathcal{B}}\ed(\hat{\varphi}^k t^j(e_k)\ed
    \mathsf{A}^j)-\hat{\varphi}^k\ed (t^j(e_k))\wedge\ed \mathsf{A}^j=0
\end{equation}
The first term in the above equation can be converted to a boundary term via
Stokes' theorem leading to, 
\begin{equation}
    \int_{\partial\mathcal{B}}\hat{\varphi}^k t^j(e_k)\ed
    \mathsf{A}^j-\int_{\mathcal{B}}\hat{\varphi}^k\ed (t^j(e_k))\wedge\ed
    \mathsf{A}^j=0
\end{equation}
Using the arbitrariness of $\hat{\varphi}^k$, we conclude that,
\begin{equation}
    \ed (t^j(e_k))\wedge\ed \mathsf{A}^j=0
\end{equation}
This is the condition  for the  critical point of the energy functional in the
direction of deformation, which is nothing but the balance of forces.  Note that
the connection 1-forms of the deformed configuration appear through the exterior
derivative of the deformed configuration's frame field.
\subsection{Stress a Lagrange multiplier}
For a hyper-elastic solid, stress is derived from a stored energy function
which may be written as a function of the deformation gradient. This assumption
permits us to write the equations of equilibrium as the Euler-Lagrange
equation of the stored energy functional. In a certain sense, the stress
generated in a hyper-elastic solid should satisfy certain integrability
condition (i.e. the existence of the stored energy function). Moreover, if
we assume the stored energy function to be translation and rotation
invariant, it implies equilibrium of forces and moments. Thus for
the hyper-elastic solid, balances of force and moment are consequences of
translation and rotation invariance; stress is only a secondary variable
introduced for writing the equations of equilibrium in a convenient way.

When formulated as a mixed problem, the stress tensor has a completely different
role. Our HW functional has deformation, deformation 1-forms and stress 2-forms
as inputs.  For a stored energy function,  viewed as a function of deformation
1-forms, translation and rotation invariance  cannot be discussed directly,
since nothing about the geometry of the co-tangent bundle from which the
deformation 1-forms were pulled back is known. In other words, there is nothing
in the stored energy function that requires the base space of the deformed
configurations co-tangent bundle to be Euclidean. The second term in
\eqref{eq:modifiedHW0} is introduced to imposes this constraint. Observe that,
in \eqref{eq:modifiedHW0}, the second term is multilinear in the input
arguments, vis-\'a-vis, stress 2-forms, differential of deformation and
deformation 1-forms. The stress $2-$form can now be thought as a Lagrange
multiplier introduced to impose the equality between differential of deformation
and deformation 1-forms. Alternatively, the equality between the differential of
deformation and deformation 1-forms implies that the deformed configuration is
Eulclidean.